\input harvmac
\def\ZZ{\hbox{Z\kern-.4emZ}}
\def\RR{\hbox{R\kern-.6emR}}
\def\ZZs{\hbox{\zfonteight Z\kern-.4emZ}}

\lref\StromingerSH{
  A.~Strominger and C.~Vafa,
  ``Microscopic Origin of the Bekenstein-Hawking Entropy,''
  Phys.\ Lett.\  B {\bf 379}, 99 (1996)
  [arXiv:hep-th/9601029].
}

\lref\CastroMS{
  A.~Castro, D.~Grumiller, F.~Larsen and R.~McNees,
  ``Holographic Description of AdS$_2$ Black Holes,''
  JHEP {\bf 0811}, 052 (2008)
  [arXiv:0809.4264 [hep-th]].
}

\lref\HartmanDQ{
  T.~Hartman and A.~Strominger,
  ``Central Charge for AdS$_2$ Quantum Gravity,''
  JHEP {\bf 0904}, 026 (2009)
  [arXiv:0803.3621 [hep-th]].
}

\lref\GuicaMU{
  M.~Guica, T.~Hartman, W.~Song and A.~Strominger,
  ``The Kerr/CFT Correspondence,''
  arXiv:0809.4266 [hep-th].
}

\lref\SenQY{
  A.~Sen,
  ``Black Hole Entropy Function, Attractors and Precision Counting of
  Microstates,''
  Gen.\ Rel.\ Grav.\  {\bf 40}, 2249 (2008)
  [arXiv:0708.1270 [hep-th]].
}

\lref\BardeenPX{
  J.~M.~Bardeen and G.~T.~Horowitz,
  ``The extreme Kerr throat geometry: A vacuum analog of AdS(2) x S(2),''
  Phys.\ Rev.\  D {\bf 60}, 104030 (1999)
  [arXiv:hep-th/9905099].
}

\lref\BalasubramanianKQ{
  V.~Balasubramanian, A.~Naqvi and J.~Simon,
  ``A multi-boundary AdS orbifold and DLCQ holography: A universal  holographic
  description of extremal black hole horizons,''
  JHEP {\bf 0408}, 023 (2004)
  [arXiv:hep-th/0311237].
}
\lref\BalasubramanianBG{
  V.~Balasubramanian, J.~de Boer, M.~M.~Sheikh-Jabbari and J.~Simon,
  ``What is a chiral 2d CFT? And what does it have to do with extremal black
  holes?,''
  arXiv:0906.3272 [hep-th].
}

\lref\BarnichBF{
  G.~Barnich and G.~Compere,
  ``Surface charge algebra in gauge theories and thermodynamic integrability,''
  J.\ Math.\ Phys.\  {\bf 49}, 042901 (2008)
  [arXiv:0708.2378 [gr-qc]].
}
\lref\CompereAZ{
  G.~Compere,
  ``Symmetries and conservation laws in Lagrangian gauge theories with
  applications to the mechanics of black holes and to gravity in three
  dimensions,''
  arXiv:0708.3153 [hep-th].
}
\lref\BarnichKQ{
  G.~Barnich and G.~Compere,
  ``Conserved charges and thermodynamics of the spinning Goedel black hole,''
  Phys.\ Rev.\ Lett.\  {\bf 95}, 031302 (2005)
  [arXiv:hep-th/0501102].
}
\lref\BanadosDA{
  M.~Banados, G.~Barnich, G.~Compere and A.~Gomberoff,
  ``Three dimensional origin of Goedel spacetimes and black holes,''
  Phys.\ Rev.\  D {\bf 73}, 044006 (2006)
  [arXiv:hep-th/0512105].
}

\lref\EmparanEN{
  R.~Emparan and A.~Maccarrone,
  ``Statistical Description of Rotating Kaluza-Klein Black Holes,''
  Phys.\ Rev.\  D {\bf 75}, 084006 (2007)
  [arXiv:hep-th/0701150].
}

\lref\DiasNJ{
  O.~J.~C.~Dias, R.~Emparan and A.~Maccarrone,
  ``Microscopic Theory of Black Hole Superradiance,''
  Phys.\ Rev.\  D {\bf 77}, 064018 (2008)
  [arXiv:0712.0791 [hep-th]].
}

\lref\BardeenPX{
  J.~M.~Bardeen and G.~T.~Horowitz,
  ``The extreme Kerr throat geometry: A vacuum analog of AdS(2) x S(2),''
  Phys.\ Rev.\  D {\bf 60}, 104030 (1999)
  [arXiv:hep-th/9905099].
}

\lref\BalasubramanianRE{
  V.~Balasubramanian and P.~Kraus,
  ``A stress tensor for anti-de Sitter gravity,''
  Commun.\ Math.\ Phys.\  {\bf 208}, 413 (1999)
  [arXiv:hep-th/9902121].
}

\lref\SkenderisWP{
  K.~Skenderis,
  ``Lecture notes on holographic renormalization,''
  Class.\ Quant.\ Grav.\  {\bf 19}, 5849 (2002)
  [arXiv:hep-th/0209067].
}

\lref\AmselEV{
  A.~J.~Amsel, G.~T.~Horowitz, D.~Marolf and M.~M.~Roberts,
  ``No Dynamics in the Extremal Kerr Throat,''
  arXiv:0906.2376 [hep-th].
}
\lref\DiasEX{
  O.~J.~C.~Dias, H.~S.~Reall and J.~E.~Santos,
  ``Kerr-CFT and gravitational perturbations,''
  arXiv:0906.2380 [hep-th].
}

\lref\MaldacenaBW{
  J.~M.~Maldacena and A.~Strominger,
  ``AdS(3) black holes and a stringy exclusion principle,''
  JHEP {\bf 9812}, 005 (1998)
  [arXiv:hep-th/9804085].
}

\lref\CveticUW{
  M.~Cvetic and F.~Larsen,
  ``General rotating black holes in string theory: Greybody factors and  event
  horizons,''
  Phys.\ Rev.\  D {\bf 56}, 4994 (1997)
  [arXiv:hep-th/9705192].
}

\lref\CveticXV{
  M.~Cvetic and F.~Larsen,
  ``Greybody factors for rotating black holes in four dimensions,''
  Nucl.\ Phys.\  B {\bf 506}, 107 (1997)
  [arXiv:hep-th/9706071].
}

\lref\CveticVP{
  M.~Cvetic and F.~Larsen,
  ``Black hole horizons and the thermodynamics of strings,''
  Nucl.\ Phys.\ Proc.\ Suppl.\  {\bf 62}, 443 (1998)
  [Nucl.\ Phys.\ Proc.\ Suppl.\  {\bf 68}, 55 (1998)]
  [arXiv:hep-th/9708090].
}
\lref\CveticAP{
  M.~Cvetic and F.~Larsen,
  ``Greybody factors for black holes in four dimensions: Particles with
  spin,''
  Phys.\ Rev.\  D {\bf 57}, 6297 (1998)
  [arXiv:hep-th/9712118].
}

\lref\LarsenGE{
  F.~Larsen,
  ``A string model of black hole microstates,''
  Phys.\ Rev.\  D {\bf 56}, 1005 (1997)
  [arXiv:hep-th/9702153].
}

\lref\DijkgraafFQ{
  R.~Dijkgraaf, J.~M.~Maldacena, G.~W.~Moore and E.~P.~Verlinde,
  ``A black hole farey tail,''
  arXiv:hep-th/0005003.
}

\lref\KrausVZ{
  P.~Kraus and F.~Larsen,
  ``Microscopic Black Hole Entropy in Theories with Higher Derivatives,''
  JHEP {\bf 0509}, 034 (2005)
  [arXiv:hep-th/0506176].
}

\lref\LarsenBU{
  F.~Larsen,
  ``Anti-de Sitter spaces and nonextreme black holes,''
  arXiv:hep-th/9806071.
}

\lref\CveticXV{
  M.~Cvetic and F.~Larsen,
  ``Greybody factors for rotating black holes in four dimensions,''
  Nucl.\ Phys.\  B {\bf 506}, 107 (1997)
  [arXiv:hep-th/9706071].
}

\lref\CveticJN{
  M.~Cvetic and F.~Larsen,
  ``Greybody Factors and Charges in Kerr/CFT,''
  JHEP {\bf 0909}, 088 (2009)
  [arXiv:0908.1136 [hep-th]].
}

\lref\MaldacenaDS{
  J.~M.~Maldacena and L.~Susskind,
  ``D-branes and Fat Black Holes,''
  Nucl.\ Phys.\  B {\bf 475}, 679 (1996)
  [arXiv:hep-th/9604042].
}
\lref\KunduriVF{
  H.~K.~Kunduri, J.~Lucietti and H.~S.~Reall,
  ``Near-horizon symmetries of extremal black holes,''
  Class.\ Quant.\ Grav.\  {\bf 24}, 4169 (2007)
  [arXiv:0705.4214 [hep-th]].
}

\lref\CveticUW{
  M.~Cvetic and F.~Larsen,
  ``General rotating black holes in string theory: Greybody factors and  event
  horizons,''
  Phys.\ Rev.\  D {\bf 56}, 4994 (1997)
  [arXiv:hep-th/9705192].
}

\lref\CveticXV{
  M.~Cvetic and F.~Larsen,
  ``Greybody factors for rotating black holes in four dimensions,''
  Nucl.\ Phys.\  B {\bf 506}, 107 (1997)
  [arXiv:hep-th/9706071].
}

\lref\CveticVP{
  M.~Cvetic and F.~Larsen,
  ``Black hole horizons and the thermodynamics of strings,''
  Nucl.\ Phys.\ Proc.\ Suppl.\  {\bf 62}, 443 (1998)
  [Nucl.\ Phys.\ Proc.\ Suppl.\  {\bf 68}, 55 (1998)]
  [arXiv:hep-th/9708090].
}

\lref\LarsenGE{
  F.~Larsen,
  ``A string model of black hole microstates,''
  Phys.\ Rev.\  D {\bf 56}, 1005 (1997)
  [arXiv:hep-th/9702153].
}

\lref\StromingerYG{
  A.~Strominger,
  ``AdS(2) quantum gravity and string theory,''
  JHEP {\bf 9901}, 007 (1999)
  [arXiv:hep-th/9809027].
}
\lref\ChoFZ{
  J.~H.~Cho, T.~Lee and G.~W.~Semenoff,
  ``Two dimensional anti-de Sitter space and discrete light cone
  quantization,''
  Phys.\ Lett.\  B {\bf 468}, 52 (1999)
  [arXiv:hep-th/9906078].
}

\lref\AzeyanagiBJ{
  T.~Azeyanagi, T.~Nishioka and T.~Takayanagi,
  ``Near Extremal Black Hole Entropy as Entanglement Entropy via AdS2/CFT1,''
  Phys.\ Rev.\  D {\bf 77}, 064005 (2008)
  [arXiv:0710.2956 [hep-th]].
}

\lref\GuptaKI{
  R.~K.~Gupta and A.~Sen,
  ``Ads(3)/CFT(2) to Ads(2)/CFT(1),''
  JHEP {\bf 0904}, 034 (2009)
  [arXiv:0806.0053 [hep-th]].
}

\lref\MaldacenaUZ{
  J.~M.~Maldacena, J.~Michelson and A.~Strominger,
  ``Anti-de Sitter fragmentation,''
  JHEP {\bf 9902}, 011 (1999)
  [arXiv:hep-th/9812073].
}

\lref\SenVZ{
  A.~Sen,
  ``Arithmetic of Quantum Entropy Function,''
  arXiv:0903.1477 [hep-th].
}

\lref\SenVM{
  A.~Sen,
  ``Quantum Entropy Function from AdS(2)/CFT(1) Correspondence,''
   arXiv: 0809.3304 [hep-th].
}

\lref\HollandsYA{
  S.~Hollands, A.~Ishibashi and D.~Marolf,
  ``Counter-term charges generate bulk symmetries,''
  Phys.\ Rev.\  D {\bf 72}, 104025 (2005)
  [arXiv:hep-th/0503105].
}

\lref\MatsuoSJ{
  Y.~Matsuo, T.~Tsukioka and C.~M.~Yoo,
  ``Another Realization of Kerr/CFT Correspondence,''
  arXiv:0907.0303 [hep-th].
}

\lref\MatsuoPG{
  Y.~Matsuo, T.~Tsukioka and C.~M.~Yoo,
  ``Yet Another Realization of Kerr/CFT Correspondence,''
  arXiv:0907.4272 [hep-th].
}
\lref\BanerjeeUK{
  N.~Banerjee, I.~Mandal and A.~Sen,
  ``Black Hole Hair Removal,''
  arXiv:0901.0359 [hep-th].
}
\lref\SenVZ{
  A.~Sen,
  ``Arithmetic of Quantum Entropy Function,''
  arXiv:0903.1477 [hep-th].
}
\lref\BredbergPV{
  I.~Bredberg, T.~Hartman, W.~Song and A.~Strominger,
  ``Black Hole Superradiance From Kerr/CFT,''
  arXiv:0907.3477 [hep-th].
}

\lref\RasmussenIX{
  J.~Rasmussen,
  ``Isometry-preserving boundary conditions in the Kerr/CFT correspondence,''
  arXiv:0908.0184 [hep-th].
}

\lref\AmselPU{
  A.~J.~Amsel, D.~Marolf and M.~M.~Roberts,
  ``On the Stress Tensor of Kerr/CFT,''
  arXiv:0907.5023 [hep-th].
}

\lref\BarnichJY{
  G.~Barnich and F.~Brandt,
  ``Covariant theory of asymptotic symmetries, conservation laws and  central
  charges,''
  Nucl.\ Phys.\  B {\bf 633}, 3 (2002)
  [arXiv:hep-th/0111246].
}

\lref\CompereIN{
  G.~Compere and S.~Detournay,
  ``Centrally extended symmetry algebra of asymptotically Goedel spacetimes,''
  JHEP {\bf 0703}, 098 (2007)
  [arXiv:hep-th/0701039].
}

\lref\CastroJF{
  A.~Castro and F.~Larsen,
  ``Near Extremal Kerr Entropy from AdS$_2$ Quantum Gravity,''
  JHEP {\bf 0912}, 037 (2009)
  [arXiv:0908.1121 [hep-th]].
}

\lref\SpradlinBN{
  M.~Spradlin and A.~Strominger,
  ``Vacuum states for AdS(2) black holes,''
  JHEP {\bf 9911}, 021 (1999)
  [arXiv:hep-th/9904143].
}
\lref\MartinecWM{
  E.~J.~Martinec,
  ``Conformal field theory, geometry, and entropy,''
  arXiv:hep-th/9809021.
}

\lref\AstorinoBJ{
  M.~Astorino, S.~Cacciatori, D.~Klemm and D.~Zanon,
  ``AdS(2) supergravity and superconformal quantum mechanics,''
  Annals Phys.\  {\bf 304}, 128 (2003)
  [arXiv:hep-th/0212096].
}

\lref\BriganteRV{
  M.~Brigante, S.~Cacciatori, D.~Klemm and D.~Zanon,
  ``The Asymptotic Dynamics of two-dimensional (anti-)de Sitter Gravity,''
  JHEP {\bf 0203}, 005 (2002)
  [arXiv:hep-th/0202073].
}

\lref\VerlindeGT{
  H.~L.~Verlinde,
  ``Superstrings on AdS(2) and superconformal matrix quantum mechanics,''
  arXiv:hep-th/0403024.
}

\lref\CadoniJA{
  M.~Cadoni and S.~Mignemi,
  ``Asymptotic symmetries of AdS(2) and conformal group in d = 1,''
  Nucl.\ Phys.\  B {\bf 557}, 165 (1999)
  [arXiv:hep-th/9902040].
}

\lref\NavarroSalasUP{
  J.~Navarro-Salas and P.~Navarro,
  ``AdS(2)/CFT(1) correspondence and near-extremal black hole entropy,''
  Nucl.\ Phys.\  B {\bf 579}, 250 (2000)
  [arXiv:hep-th/9910076].
}

\lref\EmparanPM{
  R.~Emparan, C.~V.~Johnson and R.~C.~Myers,
  ``Surface terms as counterterms in the AdS/CFT correspondence,''
  Phys.\ Rev.\  D {\bf 60}, 104001 (1999)
  [arXiv:hep-th/9903238].
}

\lref\deHaroXN{
  S.~de Haro, S.~N.~Solodukhin and K.~Skenderis,
  ``Holographic reconstruction of spacetime and renormalization in the  AdS/CFT
  correspondence,''
  Commun.\ Math.\ Phys.\  {\bf 217}, 595 (2001)
  [arXiv:hep-th/0002230].
}

\Title{
}
{\vbox{\centerline{Three Dimensional Origin of}
\medskip
\centerline{ AdS$_2$ Quantum Gravity }}}
\medskip
\centerline{\it
Alejandra Castro ${}^{\diamond}$\foot{acastro@physics.mcgill.ca}, Cynthia Keeler
${}^{\ddagger}$\foot{cakeeler@physics.harvard.edu} and Finn
Larsen${}^{\dagger}$\foot{larsenf@umich.edu}
}
\bigskip
\centerline{${}^\diamond$Physics Department, McGill University, Montreal, QC H3A 2T8, 
Canada.}
\smallskip
\centerline{${}^\ddagger$Department of Physics, Harvard University, Cambridge, MA 02138, USA.}
\smallskip
\centerline{${}^\dagger$Michigan Center for Theoretical Physics, Ann Arbor,
MI 48109, USA.}
\smallskip

\vglue .3cm
\bigskip\bigskip\bigskip
\centerline{\bf Abstract}
\noindent
We study AdS$_2$ quantum gravity with emphasis on consistency with results from
AdS$_3$. We lift AdS$_2$ black holes to three dimensions and map fluctuations around
the solutions. Comparison with near extremal BTZ are discussed,  with due emphasis on global aspects. The results confirm that parameters like central charges and conformal weights computed 
directly in 2D are consistent with standard results in 3D. 
Applying our results to the thermodynamics of near extreme Kerr black holes, we show that AdS$_2$ quantum gravity gives the correct central charge $c=12J$, and the entropy of 
excitations above the extremal limit is captured correctly.

\Date{}

\newsec{Introduction}

An illuminating way to analyze black hole dynamics is to focus on the 2D effective theory
that governs the near horizon region after dimensional reduction of the angular directions.
In extremal and near extremal settings this reasoning leads one to consider AdS$_2$
quantum gravity as a theory that encodes universal features of black holes near their
ground state \refs{\StromingerYG, \SenVM}. The expectation from holography is that this theory is dual to a 1D CFT. Indeed, AdS$_2$ quantum gravity is independently interesting as the most bare bones example of AdS/CFT correspondence.

However, many aspects of AdS$_2$ quantum gravity remain puzzling. Indeed,
propagating degrees of freedom are absent in 2D gravity quite generally. A related
fact is the existence of diffeomorphisms between AdS$_2$ black holes and global
AdS$_2$. These features indicate a particularly strong form of holography, where
all degrees of freedom are localized on the boundary. Accordingly, the dynamics
of the theory is necessarily relatively subtle, as it depends on the details of
boundary conditions satisfied by diffeomorphisms.


Our approach to AdS$_2$ quantum gravity implements relatively conventional AdS/CFT
methods (see {\it e.g.} \refs{\BalasubramanianRE, \EmparanPM,\deHaroXN, \SkenderisWP} and references within), following \refs{\CastroMS,\CastroJF}. The emphasis in this paper is to ensure that
the subtleties encountered in 2D have been addressed correctly by embedding the
theory into AdS$_3$ and establishing consistency with well-known 3D results. Locally, our embedding is valid exactly, with no limit needed.  However, in order to satisfy the correct global identifications, and recover the $SL(2,R)$ isometry of AdS$_2$ from BTZ, we take a further limit. 
We maintain excitations above the extremal ground state in this limit, in contrast
to discussions invoking a {\it strict} extremal limit of the
BTZ black hole (see {\it eg.} \refs{\StromingerYG,\SenVM,\GuptaKI,\BalasubramanianBG}).
This is significant because we interpret AdS$_2$ quantum gravity as the dynamical theory of the
{\it excitations} above the extremal limit. In contrast, the ground state degeneracy
of the extremal state is inert, because it corresponds to the direction removed
by the reduction from 3D to 2D. Other interesting approaches to AdS$_2$ quantum gravity include \refs{\CadoniJA,\NavarroSalasUP,\ChoFZ,\BriganteRV,\AstorinoBJ, \VerlindeGT, \AzeyanagiBJ}. 


Our reduction from AdS$_3$ to AdS$_2$ also extends off-shell. The restriction to fluctuations respecting asymptotically AdS$_2$ boundary
conditions imposes a constraint that removes one chiral sector from the 3D theory.
The other sector can be identified with AdS$_2$ quantum gravity. We verify that the central charges from these two points of view match. We also find that the conformal weight of the BTZ black hole 
near extremality matches that of the AdS$_2$ black hole.


A significant motivation for our work is the application of AdS$_2$ quantum gravity to
the Kerr black hole. Near extreme Kerr black holes exhibit an AdS$_2$ near horizon region,
although not one that descends from AdS$_3$. Fortunately our implementation of
AdS$_2$ quantum gravity is meaningful intrinsically in 2D --- consistency with
AdS$_3$ is ultimately for guidance and validation only. Applying our results to near
extreme Kerr black hole we recover the central charge $c=12J$ that was previously computed
using other methods \GuicaMU\ (see also \refs{\MatsuoSJ,\MatsuoPG,\RasmussenIX } ). We also determine the relation between excitation energy and
conformal weight. Taken together,
these results give an entropy of the dual CFT which is found to agree with the
area law of near extreme Kerr black holes.


It is an important feature of AdS$_3$ that symmetries {\it guarantee} agreement
between the Bekenstein-Hawking entropy of BTZ black holes and the entropy of the dual
CFT$_2$. The key observation is that BTZ black holes are related to thermal AdS$_3$ by 
a large coordinate transformation which reduces to a modular transformation on the
boundary. The spirit of our investigation is that this result should reduce to AdS$_2$
quantum gravity and we include preliminary evidence in this direction in the final 
discussion. 


This paper is organized as follows.
In section 2 we develop AdS$_2$ quantum gravity directly in 2D using standard
AdS/CFT correspondence, primarily following \refs{\CastroMS, \CastroJF}.
We review the argument that AdS$_2$ black holes constitute nontrivial states at the
quantum level \refs{\MaldacenaUZ,\SpradlinBN}.
In section 3 we embed our 2D theory into AdS$_3$ quantum gravity and show consistency
with well established local results from that setting. 
In section 4, we match the global properties of AdS$_2$ black holes to those of BTZ in 
a suitable near horizon limit. 
In section 5 we apply our results to near extreme Kerr black holes.


\newsec{AdS$_2$ Quantum Gravity}
In this section we first review some results of AdS$_2$ quantum gravity and cast
them in a robust form that can be readily compared to 3D.  In addition, we show that the non-extremal solutions are related to the vacuum via a density matrix at the boundary.

\subsec{Theory and solutions}

The 2D effective theory contains a metric $g_{\mu\nu}$ with $\mu,\nu=1,2$, 
a gauge connection ${\cal B}_\mu$ encoding rotation or electric charges of the 
higher dimensional black hole and one scalar field $\psi$ that couples to the size 
of the angular directions. 
The precise coupling among the fields in the action will depend on the details of the setting.
For definiteness we focus on the effective theory for the near extremal Kerr black hole derived 
in  \CastroJF. In this case the bulk action is
\eqn\ba{S_{\rm Kerr}={\ell^2\over 4G_4}\int
d^2x\sqrt{{-g}}\left[e^{-2\psi}{R^{(2)}}+{1\over\ell^2}
+2{\nabla}_{\mu}e^{-\psi}\,{\nabla}^{\mu}e^{-\psi}-{\ell^2\over2}e^{-4\psi}{\cal
G}^2\right]~,}
where the gauge field strength is ${\cal G}=d{\cal B}$. Effective actions for other 
black holes than Kerr will differ from \ba\ by field redefinitions and overall
normalization of the action. The discussion in the remainder of this paper applies 
to a general 2D theory of gravity with AdS$_2$ boundary conditions supported
by a Maxwell field. The field $\psi$ will be constant due to the AdS$_2$ boundary 
conditions \CastroMS. The main reference to the Kerr effective action \ba\ will be 
that expressions involve the coupling $G_4$. An alternative notation that 
exclusively refers to 2D physics amounts to the identification
\eqn\baa{
G_4 = 4\pi G_2\ell^2 e^{-2\psi}~,
}
where the 2D gravitational coupling $G_2$ in turn may be recast in terms 
of the coupling appropriate to the context at hand.

Solutions with constant $\psi$ are locally AdS$_2$ with radius $\ell$. The 
equations of motion reduce to
\eqn\bb{R^{(2)}=-e^{2\psi}{2\over\ell^2}~,\quad {\cal G}^2=-e^{2\psi}{2\over\ell^4}~, 
\quad \nabla_\mu {\cal G}^{\mu\nu}=0~.}
In a gauge where ${\cal B}_\rho=0$ the general solution takes the form
\eqn\bc{ds^2=e^{-2\psi}d\rho^2+h_{tt}dt^2~,\quad {\cal B}={\cal B}_t dt~.}
Then the black hole solutions to \bb\ are
\eqn\bca{\eqalign{
h_{tt} &= -{1\over 4} e^{-2\psi} e^{2\rho/\ell} \left(1 -
{\epsilon^2\over\ell^2}e^{-2\rho/\ell}\right)^2~,\cr
{\cal B}_t & ={1\over 2\ell} e^{\rho/\ell}\left(   1- {\epsilon\over\ell}
e^{-\rho/\ell}\right)^2~.
}}
where $\epsilon$ is the scale that controls the energy above extremality. The 
near extremal Kerr example (reviewed in Appendix A) amounts to
the identification
\eqn\bd{\eqalign{e^{-2\psi}&=1~,\cr \ell^2 &=2G_4 J~.}}

The renormalized action 
for AdS$_2$ gravity can be obtained using the standard procedure 
for AdS/CFT \refs{\SenVM, \CastroMS, \SenVZ}. The fundamental requirement
that the variational principle must be well 
defined determines the boundary action such that the on-shell action is finite. 
For \ba\ the counterterms are
\eqn\bda{S_{\rm bndy}= {\ell^2\over 2 G_4}\int
dt\sqrt{-h}\left[e^{-2\psi}K^{(2)}-{1\over
2\ell}e^{-\psi}+{\ell\over2}e^{-3\psi}{\cal
B}_a{\cal B}^a\right]~,}
with $K^{(2)}$ the extrinsic curvature at the boundary
\eqn\be{K^{(2)}={1\over 2}h^{tt}n^\mu\partial_\mu h_{tt}~,\quad 
n^{\rho}=\sqrt{g^{\rho\rho}}~. }

\subsec{Boundary conditions and asymptotic symmetries}

It is essential to note that only a combination of diffeomorphisms and gauge 
transformations are consistent with the gauge conditions imposed on the 
solutions \HartmanDQ. Here we will review the allowed transformations 
that asymptotically preserve AdS$_2$ boundary conditions.

The unspecified fields in the general solution \bc\ can be expanded 
asymptotically in the Fefferman-Graham form
\eqn\cd{\eqalign{h_{tt}&=h^{(0)}e^{2\rho/\ell}+h^{(2)}+h^{(4)}e^{-2\rho/\ell}+\cdots~,\cr{\cal 
B}_{t}&={\cal B}^{(0)}e^{\rho/\ell}+{\cal B}^{(2)}+{\cal
B}^{(4)}e^{-\rho/\ell}+\cdots~.}}
The scalar is constant and set to $\psi=0$. Solutions that are asymptotically 
AdS$_2$ satisfy the constraint
\eqn\ed{h^{(0)}+\ell^2{\cal B}^{(0)}{\cal B}^{(0)}=0~.}
This condition expresses the fact that the field strength ${\cal G}$
must be proportional to the AdS$_2$ volume two-form. On-shell we
can specify the leading behavior as
\eqn\cdb{{\cal B}^{(0)}={1\over 2\ell}~,\quad h^{(0)}=-{1\over 4}~.}
 The remaining terms in \cd, $h^{(n)}$ and ${\cal B}^{(n)}$ with $n=2,4,\ldots$, 
 are arbitrary functions of time. Notice that ${\cal B}^{(2)}$ is pure gauge, and its 
 value is fixed by demanding regularity of the solution.

Under diffeomorphisms $x^\mu\to x^{\mu}+\epsilon^{\mu}$ with parameter
\eqn\da{\eqalign{\epsilon^t &= \xi(t)-{\ell^2\over 2h^{(0)} }\partial^2_t \xi\, 
e^{-2\rho/\ell}+{\cal O}(e^{-4\rho/\ell})~,
\cr \epsilon^\rho &= -\ell\,\partial_t\xi(t)~,
}}
the gauge conditions for the metric, {\it i.e.} $g_{\rho\rho}=1$ and $g_{\rho t}=0$, 
are preserved.  However, such diffeomorphisms transform 
the gauge field such that ${\cal B}_\rho \neq 0$.
To restore the gauge condition ${\cal B}_\rho=0$ we compensate the diffeomorphism 
by a gauge transformation with parameter
\eqn\dab{\Lambda=-{\ell\over h^{(0)} }\partial^2_t \xi \left({\cal 
B}^{(0)}e^{\rho/\ell}+{1\over2}{\cal B}^{(2)}\right)\,e^{-2\rho/\ell}+{\cal
O}(e^{-3\rho/\ell})~. }

Combining the diffeomorphism \da\ and gauge transformation \dab\ preserves 
our asymptotic boundary conditions. The metric and gauge fields transform as
\eqn\dac{\eqalign {\delta_{\epsilon}h_{tt}&=\xi \partial_t h^{(2)}-2\partial_t \xi \, 
h^{(2)}-\ell^2\partial^3_t\xi +{\cal O}(e^{-2\rho/\ell})~, \cr
\delta_{\epsilon+\Lambda}{\cal B}_{t}&= \partial_t (\xi {\cal
B}^{(2)})+\left[\xi\partial_t{\cal B}^{(4)}-\partial_t \xi \, {\cal
B}^{(4)}\right]e^{-\rho/\ell}+{\ell^2{\cal B}^{(0)}\over 2h^{(0)}}\partial^3_t\xi \,
e^{-\rho/\ell}+ {\cal O}(e^{-2\rho/\ell})~.}}

\subsec{Conserved charges}

There are two relevant transformations in our discussion, diffeomorphisms and gauge 
transformations. For each of them there is a corresponding generator (a boundary 
current) which we want to construct.
\vskip 0.2cm
\noindent {\bf Generator of diffeomorphisms}.  The variation of the action under a 
diffeomorphism is given by
\eqn\ea{\delta_{\epsilon}S=\int dt \sqrt{-h} \, \left[ {1\over 2}T^{ab}\delta_\epsilon h_{ab}+ 
 \sqrt{-h} J^{a}\delta_\epsilon {\cal B}_{a}\right] +{\rm (e.o.m)}~,}
where
 \eqn\eb{\eqalign{\delta_\epsilon h_{ab}&=\nabla_a \epsilon_b + \nabla_b 
 \epsilon_a~,\cr
\delta_\epsilon {\cal B}_a&=\epsilon^b\nabla_b {\cal B}_a +{\cal B}_b\nabla_a
\epsilon^b~,}}
and
\eqn\ec{\eqalign{T_{tt}&=-{\ell^2\over 4 G_4}\left(
{1\over\ell}e^{-\psi}h_{tt}+\ell e^{-3\psi}{\cal B}_t{\cal
B}_t\right)~,\cr J_t&= {\ell^2\over 2
G_4}e^{-3\psi}\left(-e^{-\psi}\ell^2n^\mu{\cal G}_{\mu
t}+\ell{\cal B}_t\right)~.
}}
Inserting the Fefferman-Graham expansion \cd\ these currents become 
\eqn\ega{\eqalign{T_{tt} &=-{\ell\over 4G_4}\left[2\ell^2 
{\cal B}^{(0)}{\cal B}^{(2)}+(h^{(2)}+\ell^2
{\cal B}^{(2)}{\cal B}^{(2)}+2\ell^2{\cal B}^{(0)}{\cal B}^{(4)})e^{-\rho/\ell}+{\cal
O}(e^{-2\rho/\ell})\right]e^{\rho/\ell}~,\cr
J_t&=
{\ell^2\over2G_4}\left[{\cal B}^{(2)}+2{\cal B}^{(4)}e^{-\rho/\ell}+{\cal O}(e^{-2\rho/\ell})\right]
~.}}

Usually the generator of diffeomorphisms is just the stress tensor $T_{ab}$
because the matter terms in \ea\ decay more rapidly than the metric field. In 
2D the situation is different since the rapid increase ${\cal B}_t\sim e^{\rho/\ell}$ 
at the boundary is needed to support the solution. This situation imposes 
the constraint \ed\ between the leading terms in the Fefferman-Graham expansion
\cd, which in turn will imply that the two terms in \ea\ are of the same order. 
Explicitly, to preserve the asymptotic AdS$_2$ boundary conditions the variations of
metric and gauge field must be related as
\eqn\ef{\delta h^{(0)}=-2\ell^2{\cal B}^{(0)}\delta{\cal B}^{(0)}~. }
so that \ea\ becomes
\eqn\eg{\eqalign{
\delta_{\epsilon}S &=\int dt \sqrt{-h} \left(- \ell^2 {\cal B}^{(0)}T^{tt} 
e^{\rho/\ell} + J^{t}\right)e^{\rho/\ell}\delta {\cal B}^{(0)} \cr
&={2\ell^2\over G_4}\int dt \sqrt{-h} 
\left[h^{(2)}+\ell^2{\cal B}^{(2)}{\cal B}^{(2)}-2\ell^2{\cal B}^{(0)}{\cal B}^{(4)}
+{\cal O}(e^{-\rho/\ell})\right] e^{-2\rho/\ell}\delta {\cal B}^{(0)} \cr
&={2\ell^2\over G_4}\int dt \sqrt{-h}\, \ell^2{\cal 
B}^{(2)}{\cal B}^{(2)} \,e^{-2\rho/\ell}\delta {\cal B}^{(0)}~.}}
In the second line we used the currents \ega\ and the on-shell values \cdb. 
In the last line we used the value $h^{(2)} = 2\ell^2{\cal B}^{(0)}{\cal B}^{(4)}$
imposed by the equations of motion, whose perturbative expansion is detailed 
in Appendix B.

The time translation operator is usually interpreted as the energy of the configuration. 
Accordingly, \eg\ assigns an energy of the form $E\sim{\cal B}^{(2)2}$ to 
excitations in the theory. To avoid unnecessary dependence on conventions
we will not specify the proportionality constant in the energy, but just work with 
variations of the action. We are especially interested in the specific case of the AdS$_2$ 
black hole \bca\ for which
\eqn\eha{\eqalign{\delta_{\epsilon}S&
= {2\over G_4}\int dt \sqrt{-h}\, \epsilon^2 \,e^{-2\rho/\ell}\delta {\cal B}^{(0)}~.}}
In the next section we will recover this generator from the 3D perspective. 

\vskip 0.2cm
\noindent {\bf Generator of gauge transformations}.  The variation of the action under 
a gauge transformation is
\eqn\fa{\delta_{\Lambda}S= \int dt \sqrt{-h} {\cal J}^{a}\delta_\Lambda {\cal 
B}_{a}~,}
where
\eqn\fb{{\cal J}_t= {\ell^3\over 2
G_4}e^{-3\psi}{\cal B}_t ~.
}
Allowed gauge transformations correspond to variations of 
the action with respect to the constant term in the gauge field ${\cal B}^{(2)}$ and 
not ${\cal B}^{(0)}$, the leading term \foot{We specify the gauge ${\cal B}_\rho=0$ so 
residual gauge transformations are independent of $\rho$, but they may depend
on the boundary coordinate $t$.}.  Using our asymptotic expansion \cd, the 
generator \fa\ reads
\eqn\fc{\eqalign{\delta_{\Lambda}S&= -{2\ell^3\over G_4}\int dt \sqrt{-h}\, {\cal 
B}^{(0)}e^{-\rho/\ell}\delta_\Lambda {\cal B}^{(2)}\cr &
=-{\ell^2\over G_4}\int dt \sqrt{-h}\, e^{-\rho/\ell}\delta_\Lambda {\cal B}^{(2)}~.}}
Since $\sqrt{-h}\sim e^{\rho/\ell}$, the allowed gauge transformations modify the action 
by at most a constant term \CastroMS. 

The generator of gauge transformations \fc\ can be interpreted as the charge of the 
electric field that supports the solution \CastroJF. Our interest will be in comparing
with the corresponding 3D result and for that purpose the variation \fc\ is
a convenient and unambiguous expression for the charge.\foot{In situations where
the gauge transformation is accompanied by a diffeomorphism the transformation of
${\cal B}^{(2)}$ is dominated by the gauge current 
${\cal J}_t\sim {\cal O}(e^{\rho/\ell})$ while $J^t\sim {\cal O}(1)$.}

\subsec{Thermal states in AdS$_2$}

An important issue to address is if the solutions \bc\ - \bca\  to the 2D theory correspond to distinct states of the dual theory or if they are trivially related by a diffeomorphism. To explore this question we will see how an AdS$_2$ black hole ($\epsilon \neq 0$) is related to the vacuum solution ($\epsilon=0$). It is convenient to write \bc\ as
\eqn\pb{ds^2=-{U^2-\epsilon^2\over \ell^2}dt^2+{\ell^2\over U^2-\epsilon^2}dU^2~, \quad {\cal B}={U-\epsilon\over\ell^2}dt~.}
where the coordinate $U$ is related to the radial coordinate in \bca\ as
\eqn\pba{
U = {\ell\over 2} e^{\rho/\ell}\left( 1 + {\epsilon^2\over\ell^2}e^{-2\rho/\ell}\right)~.
}
The value of the modified stress tensor \eg\ evaluated on \pb\ is
\eqn\pbd{T_{tt} + {\cal B}_t J_t = {\ell^3\over 4G_4} {\cal B}^{(2)2} = {\ell\over 4G_4}  {\epsilon^2\over\ell^2}~.}

The vacuum solution is AdS$_2$ in Poincare coordinates
\eqn\pc{ds^2={\ell^2\over y^2}(dy^2-dw^2)~,\quad {\cal B}=-{\ell\over y}dw~.}
The charge associated to diffeomorphisms for this background is zero, {\it i.e.} $T_{tt} + {\cal B}_t J_t =0$. 

The transformation that relates the black hole \pb\ and \pc\  is given by
the coordinate change
\eqn\pca{\eqalign{y={\ell \epsilon\over\sqrt{U^2-\epsilon^2}} e^{-\epsilon t/\ell^2}~,\cr w=\ell {U\over\sqrt{U^2-\epsilon^2}} e^{-\epsilon t/\ell^2}~, }}
accompanied by the gauge transformation
\eqn\pcaa{\Lambda=-{1\over2}\ln\left(U+\epsilon\over U-\epsilon\right)~.}
There are a couple of things worth mentioning here. First, the coordinate transformation \pca\ is singular at the horizon $U=\epsilon$. Second, near the boundary  $U\to \infty$ we have
\eqn\pcb{w=\ell \, e^{-\epsilon t/\ell^2} +{\cal O}(U^{-1})~.}
This relation mimics that between Rindler and Minkowski coordinates.  In that
case, the quantum states are related by a density matrix such that the Rindler
observer will detect a thermal bath with temperature
\eqn\pdd{T={\epsilon\over 2\pi \ell^2}~.}

The modified boundary stress tensor $T_{tt} + {\cal B}_t J_t$ transforms as an affine 
tensor in the usual manner so the transformation \pcb\ induces the anomalous 
energy 
\eqn\pd{\eqalign{T_{tt} + {\cal B}_t J_t &={c\over12}\ell\{w,t\}(\partial_wt)^{-2}\cr &= {\pi^2 \over 6}c\,\ell\,\,T^2~,}}
where the Schwarzian derivative
\eqn\pda{\{w,t\}={\partial_w^3t\over \partial_w t}-{3\over2}\left({\partial_w^2t\over \partial_w t}\right)^2={1\over 2w^2}~.}
Comparing with \pbd\ found by evaluation on the black hole solution we deduce the
central charge 
\eqn\pdc{c={6\ell^2\over G_4}~.}
This is the same result found in \CastroJF, by examination of the transformation properties
of the modified boundary stress tensor. 

Note that in \pd\ we are measuring energies with respect to the AdS$_2$ radius $\ell$. This scale is a natural one to use in the 2D theory, but arbitrary since the theory is exactly conformal. In section 5, when comparing with the Kerr black hole (which breaks conformal invariance), we will compute the scale $R$ and verify that energies near extremality agree after an appropriate rescaling. 

In summary, we have exhibited a classical diffeomorphism that relates the non-extremal 
AdS$_2$ metric $(\epsilon\neq 0)$ to the extremal one. In the quantum theory the
transformation at the boundary induces a thermal state due to a scale anomaly. 
Thus the two AdS$_2$ metrics correspond to different quantum states.


\newsec{AdS$_2$ Quantum Gravity from 3D Gravity}

In this section we provide a map between 2D solutions and the 3D configurations. We show that the boundary currents in 3D reproduce the same
values as those found in 2D. Finally, we lift the asymptotic symmetry group in 2D and find the effects on the 3D stress tensor. This will allow us to identify the central charges in 3D and 2D.

\subsec{Reduction of the stress tensor}
We start by  showing how the generators of diffeomorphisms and charges in AdS$_2$
quantum gravity can be recovered from the 3D stress tensor. The starting point is three dimensional gravity described by the action
\eqn\ga{I={1\over 16\pi G_3}\int d^3x \sqrt{-g}\left(R^{(3)}+{2\over L^2}\right)
+{1\over 8\pi G_3}\int d^2y \sqrt{-\gamma}\left(K^{(3)}-{1\over L^2}\right)~,}
where we have included the bulk and boundary actions. The stress tensor $T^{ab}_{\rm 3D}$ is defined via
\eqn\gb{\delta I={1\over 2}\int d^2y \sqrt{-\gamma}T^{ab}_{\rm 3D}\delta \gamma_{ab}~.}
This definition gives
\eqn\cf{T_{ab}^{\rm 3D}={1\over 8\pi G_3}\left(K_{ab}^{(3)}-K^{(3)}\gamma_{ab}+{1\over
L}\gamma_{ab}\right)~,}
where $K_{ab}^{(3)}$ is the extrinsic curvature.
%
%
Here $y^a$ with $a=1,2$ are the boundary coordinates, and the metric $\gamma_{ab}$ is
defined as
\eqn\hab{ds^2_3=L^2d\eta^2+\gamma_{ab}dy^ady^b~.}
In the 3D gravity, the boundary metric has a Fefferman-Graham
expansion of the form
\eqn\hcd{\gamma_{ab}=e^{2\eta}\gamma_{ab}^{(0)}+\gamma_{ab}^{(2)}
+e^{-2\eta}\gamma_{ab}^{(4)}+ \cdots~, }
where the dots correspond to higher powers of $e^{-2\eta}$.

We need to cast \gb\ as a variation with respect to the 2D fields $h_{tt}$ and
${\cal B}_t$ instead of $\gamma_{ab}$. This will allow us to express the 2D currents
discussed in the previous section in terms of $T^{\rm 3D}_{ab}$. We will use the 
Kaluza-Klein reduction from 3D to 2D, given by
\eqn\hc{ds^2=\ell^2(d\theta +{\cal B}_tdt)^2+d\rho^2+h_{tt}dt^2~.}
Inserting the 2D black holes \bca, we find that the 3D lift \hc\ becomes
\eqn\xc{
ds^2_{3} =d\rho^2 + \ell e^{\rho/\ell}\left(   1- {\epsilon\over\ell}
e^{-\rho/\ell}\right)^2\left[dtd\theta-{\epsilon\over \ell^2}dt^2\right] +\ell^2 d\theta^2~.}
We would like to rewrite this metric in coordinates natural from the 3D point of view; that is, we want to rewrite \xc\ in the form \hab.  Comparing \xc\ with \hab\ and \hcd\ we get 
\eqn\xxca{\rho=L \eta~, \quad 2\ell=L~.}
We can continue to use $(t,\theta)$ to describe the remaining two dimensions, but eventually we would like to compare to light-like coordinates natural for 3D boundary. If we describe the boundary of AdS$_3$ as
\eqn\xf{\gamma_{ab}^{(0)}dx^adx^b=L^2dx^+dx^-~,}
then from the leading term in \xc\ we find
\eqn\xxcb{\eqalign{x^-=&{t\over 2\ell}~,\cr 
x^+=&{\theta\over 2}-{\epsilon\over2 \ell^2}t~.}}

For now, we will continue in the combined coordinates $(\eta,t,\theta)$.  Additionally, for sake of simplicity, throughout this section we set the dilaton to the on-shell
value $\psi=0$. We now write $\gamma_{ab}$ in terms of $h_{tt}$ and ${\cal B}_t$.  From \hc\ the boundary metric in $(t,\theta)$ coordinates is
\eqn\hcb{\gamma_{tt}=h_{tt}+\ell^2{\cal B}_t{\cal B}_t~, \quad \gamma_{t\theta}=
\ell^2 {\cal B}_t~, \quad \gamma_{\theta\theta}=\ell^2~.}

Using the 2D Fefferman-Graham expansion \cd\ in \hcb\ we get
\eqn\hdb{\eqalign{\gamma_{tt}&=\left[h^{(0)}+\ell^2{\cal B}^{(0)}{\cal
B}^{(0)}\right]e^{4\eta} + 2\ell^2{\cal B}^{(2)}{\cal B}^{(0)}e^{2\eta} \cr
&\quad+\left[h^{(2)}+\ell^2{\cal B}^{(2)}{\cal B}^{(2)}+2\ell^2{\cal B}^{(0)} {\cal
B}^{(4)}\right]+\cdots~,\cr
\gamma_{t\theta}&=\ell^2{\cal B}^{(0)}e^{2\eta} + \ell^2{\cal B}^{(2)} +\ell^2{\cal
B}^{(4)} e^{-2\eta}+\cdots~, \cr \gamma_{\theta\theta}&=\ell^2~,}}
where we used $\rho/\ell=2\eta$ according to \xxca.

Before performing the reduction, we should be more precise about how to manipulate and
interpret the variation of the action \gb\ and the resultant stress tensor \cf.  In \hcd\  $\gamma^{(0)}$
is the boundary metric up to a conformal transformation and the sub-leading terms
contain information about the mass and angular momentum of the solution. According to
the AdS/CFT dictionary, the stress tensor is obtained via the variation of the action
with respect to the boundary metric
\eqn\hcda{\delta\gamma_{ab}=e^{2\eta}\delta\gamma_{ab}^{(0)}~. }
Thus, it is the variation of the action with respect to $\gamma^{(0)}$ which we will be
interested in comparing to the 2D results.  Of course, the stress tensor itself will still be as given in \cf;
we need only make the replacement \hcda\ when writing the expression for the variation of the action.

There is an additional subtlety which we must address.
At first glance, the metric in \hdb\ indicates leading behavior $\gamma_{tt}\sim e^{4\eta}$
which is not the correct asymptotic behavior. However, we restrict to
asymptotically AdS$_2$ geometry by imposing the constraint \ed\
which forces the first term in \hdb\ for
$\gamma_{tt}$ to vanish. Thus the variation of the 3D boundary metric \hcda\  by
changes  relevant on the 2D boundary is
\eqn\fb{\eqalign{\delta\gamma_{tt}&=2\ell^2{\cal B}^{(2)}e^{2\eta}\delta{\cal
B}^{(0)}+ 2\ell^2{\cal B}^{(0)}e^{2\eta}\delta {\cal B}^{(2)}~,\cr
\delta \gamma_{t\theta}&=\ell^2e^{2\eta}\delta {\cal B}^{(0)} ~,\cr \delta
\gamma_{\theta\theta}&=0~.}}
We re-write the variation of the 3D
action \gb\ using the $\gamma^{(0)}$ component of \fb\ which gives
\eqn\fbaa{\eqalign{\delta I=&\int d^2y\sqrt{-h}\,\ell^2e^{2\eta}\left[{\cal
B}^{(2)}T^{tt}_{\rm 3D}+ T^{t\theta}_{\rm 3D}\right]\delta {\cal B}^{(0)} +\int
d^2y\sqrt{-h}\,\ell^2e^{2\eta}\left[{\cal B}^{(0)}T^{tt}_{\rm 3D}\right]\delta{\cal
B}^{(2)}~.}}

The next step is to compute the 3D stress tensor  given by \cf\ as a function of 2D fields $h^{(n)}$ and ${\cal B}^{(n)}$. The details of the calculation are presented in appendix B, and the final result for the components of $T_{ab}^{\rm 3D}$ with the indices
raised is
\eqn\fef{\eqalign{T_{\rm 3D}^{tt}&={1\over 8\pi G_3 L}{e^{-4\eta} \over \ell^2 {\cal
B}^{(0)}{\cal B}^{(0)}}\left[1+2e^{-4\eta}{{\cal B}^{(4)}\over {\cal B}^{(0)}}+{\cal
O}(e^{-6\eta})\right]~,\cr T_{\rm 3D}^{t\theta}&=-{1\over 8\pi G_3 L}{e^{-4\eta} \over
\ell^2{\cal B}^{(0)}{\cal B}^{(0)}}\left[{\cal B}^{(2)}+2e^{-2\eta}{\cal
B}^{(4)}+{\cal O}(e^{-4\eta})\right].}}
Separating the contributions due to $\delta{\cal B}^{(0)}$ and $\delta{\cal B}^{(2)}$ in
\fbaa, we find
\eqn\fbx{\eqalign{\delta I_{(0)}\equiv&\int d^2y\sqrt{-h}\,\ell^2e^{2\eta}\left[{\cal
B}^{(2)}T^{tt}_{\rm 3D}+ T^{t\theta}_{\rm 3D}\right]\delta {\cal B}^{(0)} \cr
=&- {1\over G_3\ell }\int dt\sqrt{-h}\, \epsilon^2 e^{-4\eta}\delta {\cal B}^{(0)}~,
 }}
and
\eqn\px{\eqalign{
\delta I_{(2)}\equiv &\int d^2y\sqrt{-h}\,\ell^2e^{2\eta}\left[{\cal
B}^{(0)}T^{tt}_{\rm 3D}\right]\delta {\cal B}^{(2)} \cr =&{\ell\over 2G_3}\int
dt\sqrt{-h}\, e^{-2\eta}\delta {\cal B}^{(2)}~,}}
where we have used (B.5) and
\eqn\pxa{\int d^2y=2\pi L  \int dt~.}

The $\delta I_{(0)}$ and $\delta I_{(2)}$ in these expressions should be compared with
the variations of the 2D action computed in section 2.2. We find that \fbx\ agrees precisely
with the generator of diffeomorphisms \eha, and \px\ with
the generator of gauge transformations \fc:
\eqn\py{\eqalign{\delta I_{(0)} = -\delta S_\epsilon~,\quad  \delta I_{(2)} = -\delta
S_\Lambda}~,}
after the identification
\eqn\pya{ G_4= 2 \ell G_3~,}
and taking into account the identification in \xxca.

As we have mentioned previously, our choice to normalize the 2D action using the 4D
coupling $G_4$ is to facilitate comparison with Kerr. However, the identification
of couplings in \pya\ clearly applies in other contexts as well, after introducing the 2D
coupling through \baa.

The map constructed in the beginning of this section determines the identification between
solutions of 3D and 2D theories. The agreement \py\ goes further, by showing an
an exact match of the boundary currents in the two theories.


\subsec{Lift of 2D asymptotic symmetry group to 3D }

In section 2.1 we constructed the asymptotic symmetry group in 2D, which includes
diffeomorphisms \da\ and gauge transformations \dab.  Here we lift these
transformations with the purpose of understanding the effects of 2D asymptotic
symmetries from a 3D point of view.

Start by considering the three dimensional line element
\eqn\ma{ds^2_{3}  = \ell^2 \left(d\theta + {\cal B}\right)^2 + d\rho^2+h_{tt}dt^2~.}
A 2D diffeomorphism acting on $(\rho,t)$  
\eqn\mba{\eqalign{t&\to t +\epsilon^t~, \cr \rho &\to \rho +\epsilon^\rho~,}}
also corresponds to a diffeomorphism in 3D.
On the other hand,  a gauge transformation ${\cal B}\to {\cal B} +d\Lambda$ lifts to a
diffeomorphism on the KK  direction
\eqn\mbb{ \theta \to \theta +\Lambda~. }
Using in \mba\ and \mbb\  the 2D parameters \da\ and \dab, the components of \ma\
transform as
\eqn\pbc{\eqalign{d\theta +{\cal B}_t dt &\to d\theta +{\cal B}_t dt+{\ell^2{\cal
B}^{(0)}\over 2h^{(0)}}\partial^3_t\xi\, e^{-\rho/\ell}dt +  {\cal
O}(e^{-2\rho/\ell})~,\cr  h_{tt}dt^2 &\to h_{tt} dt^2-{\ell^2}\partial^3_t\xi  dt^2
+2{\ell}\partial^2_t\xi d\rho dt+ {\cal O}(e^{-2\rho/\ell})~, \cr d\rho^2&\to d\rho^2
-2{\ell}\partial^2_t\xi d\rho dt~, }}
hence the line element transforms as
\eqn\pc{ds^2_3\to ds^2_3+\left({\ell^2{\cal B}^{(0)}{\cal B}^{(0)} \over h^{(0)} }
-1\right){\ell^2}\partial^3_t\xi  dt^2 =ds^2_3-2{\ell^2}\partial^3_t\xi  dt^2 ~,}
where we applied the constraint \cdb\ that imposes asymptotic AdS$_2$ boundary conditions.
Thus we see that 2D diffeomorphisms generate an anomalous transformation on the 3D metric.
Indeed, the additional term is exactly the anomaly for the $T_{tt}^{\rm 3D}$ component of stress
tensor in 3D.  Now working in the $x^\pm$ coordinates defined in \xxcb, we have
\eqn\pcb{t=L x^-~,}
and so the anomalous transformation would be
\eqn\pcc{ds^2_3\to ds^2_3-2{\ell^2}\partial^3_-\xi^-  (dx^-)^2~,}
where $\xi^-=L\xi(t)$. Using \cf\ to compute the stress tensor for \pcc\ we have
\eqn\dcd{\eqalign{T_{--}&=-{1\over 8\pi G_3 L}(2\ell^2)\partial^3_-\xi^-\cr &=
-{\ell\over 8\pi G_3 }\partial^3_-\xi^-\cr &=-{c\over 24\pi }\partial^3_-\xi^-~.}}
In the last line we introduced the normalization of \BalasubramanianRE\ for the central charge.  It implies
\eqn\dce{c={3\ell\over  G_3 }={3L\over  2G_3}~,}
the Brown-Henneaux central charge. It is an important consistency check that this agrees 
with our 2D result for the central charge \pdc, due to the identification \pya. 

Recall that the isometry group of AdS$_2$ is $SL(2,R)_R\times U(1)_L$ from the
3D point of view. Because the asymptotic symmetries in 2D only affect the $x^-$ 
component of the stress tensor, our result \dce\ is the right moving central charge $c=c_R$
due to an enhancement of $SL(2,R)_R$ symmetries to a Virasoro algebra. The 
central charge computed for extremal black holes 
in {\it e.g.} \refs{ \GuicaMU,\BalasubramanianBG} considers an enhancement of $U(1)_L$ 
to a Virasoro algebra with central charge $c_L$. That central charge must be the same
as ours, since the 3D theory is not chiral, but its origin and interpretation is different. 


\newsec{Global Aspects of the Embedding into 3D}
So far we have focused on local properties of quantum gravity.  We have lifted
solutions of AdS$_2$ gravity to those of AdS$_3$.  However we would like to do
more; we want to identify the AdS$_2$ black hole (which is locally AdS$_2$)
with the BTZ black hole (which is locally AdS$_3$).  In order to do so, we will
have to discuss the {\it global} aspects of the embedding; we will do so in
this section

\subsec{The AdS$_2$ black hole and the BTZ solution}

The BTZ black hole in Schwarzschild coordinates is
\eqn\za{ds^2={L^2r^2\over (r^2-r^2_+)(r^2-r^2_-)}dr^2-{ (r^2-r^2_+)(r^2-r^2_-)\over 
r^2}dT^2+r^2\left(d\phi+{r_+r_-\over r^2}dT\right)^2 ~,}
where $L$ is the AdS$_3$ radius and $r_{\pm}$ are the outer and inner horizon 
coordinates. For nonextremal black holes $r_+\neq r_-$ we can make the 
Fefferman-Graham (FG) expansion manifest by using the coordinates
\eqn\zb{r^2=r_+^2\cosh^2(\eta-\eta_0)-r_-^2\sinh^2(\eta-\eta_0)~,}
%
\eqn\zca{w^{\pm}=\phi\pm T~,}
with
\eqn\zba{e^{2\eta_0}={r_+^2-r_-^2\over 4L^2}~.}
Inserting in \za\ we find the BTZ black hole in the FG form
\eqn\zcb{\eqalign{ds^2=&L^2d\eta^2+\left(L^2e^{2\eta}+{1\over 
16L^2}(r_+^2-r_-^2)^2e^{-2\eta}\right)dw^+dw^-\cr &+{1\over
4}(r_+-r_-)^2(dw^-)^2+{1\over 4}(r_++r_-)^2(dw^+)^2}~.}

We can compare this expression with the lifted 2D black hole solution \xc, 
which is similarly written in FG form. For this we rescale 
the radial coordinate as in \xxca\ and transform the boundary coordinates
linearly as 
\eqn\xdc{\eqalign{w^-&={2\epsilon\over \ell (r_+-r_-)}t ={ r_++r_-\over 
8\ell^2}t ~,\cr w^+&= {2\over r_++r_-}\left( \ell\theta-{\epsilon\over \ell}t\right)= {2\ell\over r_++r_-}\theta
- {r_+-r_-\over 8\ell^2}t ~.}}
The parameters of the solutions are identified as 
\eqn\xcc{L=2\ell~,\quad {\epsilon \over \ell}={r_+^2-r_-^2\over 16 \ell^2}~.}

The 2D temporal coordinate $t$ is identified with $w^-$ for all the BTZ black holes, whether 
extremal or not. 
The complementary coordinate $\theta$, parametrizing the KK circle,
approaches the null coordinate $w^+$ in the extremal limit $r_+\to r_-$. 
The parameter $\epsilon$ describes the departure from the extremal limit, ie.
the degree of non-nullness in the dimensional reduction. 

The map \xdc-\xcc\ identifies the AdS$_2$ black hole with BTZ {\it locally},
but it also exhibits a {\it global} discrepancy. Indeed, the BTZ is defined with the identification 
$\phi \sim \phi +2\pi$, corresponding to simultaneous shift of $w^+$ and $w^-$
by $2\pi$. On the other hand, the KK lifted AdS$_2$ black hole has 
\eqn\scb{ \theta \sim \theta+4\pi~,}
because the length of the KK  circle is $2\pi L=4\pi\ell$. Comparing with 
\xdc\ we see that the periodicities of the two solutions differ, as claimed.

Despite the general discrepancy, there is a near horizon limit where the 
identifications do coincide, as discussed in detail in  \BalasubramanianBG. 
The DLCQ limit is implemented by changing the coordinates as
\eqn\wa{w^- \to {w^-\over \lambda}~, \quad e^{2\eta}\to \lambda e^{2\eta}~,}
and writing the parameters of the solution as
\eqn\wb{r_+-r_-= 4\lambda \epsilon~.}
After those replacements, simply take $\lambda\to 0$ with all other quantities fixed. 
In this limit $w^-$ has been unwrapped so that it is no longer periodic, consistent with its 
identification as 2D time in the first 
equation in \xdc. The $2\pi$ periodicity of $w^+$ remains after the limit so 
the second equation in \xdc\ imposes
\eqn\wba{
r_+ + r_- = 4\ell~,
}
for consistency with \scb.

After the DLCQ limiting procedure the solution \zcb\ reads
\eqn\wbc{\eqalign{ds^2=&L^2d\eta^2+\left(L^2e^{2\eta}+
4\epsilon^2 e^{-2\eta}\right)dw^+dw^-
+4\epsilon^2(dw^-)^2+L^2 (dw^+)^2~,}}
and the coordinate identifications \xdc\ simplify to \xxcb, now in the form
\eqn\fla{\eqalign{
w^- & = {1\over 2\ell} t \cr
w^+ & =  {1\over 2}\theta - {\epsilon\over 2\ell^2} t~.
}}
The metric \wbc\ is superficially similar to \zcb\ but it differs in two crucial aspects. First, as we
have emphasized, the scaling \wa\ unwraps the compact coordinate $w^-$ in \zcb\ 
so that it becomes non-compact in \wbc. Thus \wbc\ with \fla\ inserted is completely equivalent
to the lifted AdS$_2$ black hole \xc, including global aspects. Next, it is evident 
from \wbc\ that a finite energy-momentum tensor should be assigned to excitations with 
finite $\epsilon$ {\it after} taking the limit. In other words, whereas the original solution \zcb, 
as well as the parameter $\epsilon$ in \xcc, might give the impression that the energy of excitations 
vanish in the extremal limit \wb, the energy is in fact finite due to the simultaneous 
coordinate rescalings \wa. 

In summary, after the DLCQ limit required by the global aspects, the mass of the original 
3D black hole is fixed by \wba. The limiting theory assigns finite energy to the excitations 
above the extremal limit parametrized by $\epsilon$. 

\subsec{AdS$_2$ black hole as a quotient}
It is illuminating to implement the DLCQ limit in the construction of the BTZ
black hole as a quotient of AdS$_3$. 

In the region outside the horizon, the coordinate change
\eqn\baz{\eqalign{
z^+ & = \sqrt{r^2-r^2_+\over r^2-r^2_-} e^{2\pi T_+ (\phi+T)}~,\cr
z^- & = \sqrt{r^2-r^2_+\over r^2-r^2_-} e^{2\pi T_- (\phi-T)}~,\cr
y & = \sqrt{r^2_+ - r^2_-\over r^2 - r^2_-} e^{\pi T_+ (\phi+T)+\pi T_- (\phi-T)}~,
}}
with the temperatures
\eqn\bbz{\eqalign{
T_+ & = {r_+ + r_-\over 2\pi L}~,\cr
T_- & = {r_+ - r_-\over 2\pi L}~,
}}
maps the BTZ black hole \za\ to Poincar\'{e} AdS$_3$
\eqn\bc{
ds^2_3  = {L^2\over y^2} ( dy^2 + dz^+ dz^-)~.
}
The $2\pi$ identification on $\phi$ therefore amounts to simultaneous 
identifications of the  Poincar\'{e} coordinates
\eqn\sdb{\eqalign{
z^+ & \sim z^+ e^{4\pi^2 T_+}~,\cr
z^- & \sim z^-e^{4\pi^2 T_-}~,\cr
y & \sim y\, e^{2\pi^2 (T_-+T_+)} ~.
}}
These exponential identifications correspond to the hyperbolic/hyperbolic conjugacy class
in the $SL(2,R)_L\times SL(2,R)_R$ symmetry of AdS$_3$.

The analogous classification for the AdS$_2$ black hole follows by simplifying the
temperatures \bbz\ using \wb, \wba\ and then employing the coordinates \fla. The
embedding \baz\ into Poincar\'{e} AdS$_3$ becomes
\eqn\baz{\eqalign{
z^+ & = \tanh(\eta-\eta_0) e^{\theta- \epsilon t/\ell^2}~,\cr
z^- & =  \tanh(\eta-\eta_0) e^{\epsilon t/\ell^2}~,\cr
y & = {1\over\cosh(\eta-\eta_0)} e^{\theta/2}~.
}}
The $4\pi$ identification on $\theta$ \scb\ therefore amounts to simultaneous 
identifications of the  Poincar\'{e} coordinates
\eqn\sdc{\eqalign{
z^+ & \sim z^+ e^{4\pi}~,\cr
y & \sim y\, e^{2\pi} ~.
}}
These identifications correspond to hyperbolic/identity conjugacy class
in the $SL(2,R)_L\times SL(2,R)_R$ symmetry of AdS$_3$.

That the $z^-$ coordinate does not participate in the identifications is due to 
the noncompactness of the $R$ sector. In terms of symmetries, the hyperbolic 
identification in the $z^+$ coordinate breaks $SL(2,R)_L\to U(1)_L$, as it 
does for BTZ. However, the $SL(2,R)_R$ symmetry remains unbroken for the 
AdS$_2$ black hole, since there is no identification of $z^-$. The preserved 
symmetry is the isometry of the AdS$_2$ geometry \BalasubramanianBG. 

It is instructive to compare the DLCQ limit with the usual extremal limit, 
where one takes $T_-\to 0$ without rescaling $w^-$ as in \wa. In this case 
$w^-$ remains periodic, and so there is no restoration of $SL(2,R)_R$. The map to Poincar\'{e} \baz\ degenerates in this extreme limit but a modified 
map (with $z^-$ shifted and rescaled) shows that the conjugacy class is elliptic/parabolic.
The parabolic identification in the R-sector indeed breaks $SL(2,R)_R\to U(1)_R$.  
Regular particle states are elliptic on both sides \refs{\MaldacenaBW,\MartinecWM},
and so they similarly break both $SL(2,R)$'s. 

\subsec{The energy revisited}
It is worth revisiting our notion of energy in AdS$_2$ in view of our discussion of 
global identifications in the reduction from 3D. 

As we have argued, the dynamical theory is tied to the $R$ sector. The relation \baz\
between $z^-$ and 
the 2D time $t$ implies the 2D temperature $T_{\rm phys} = {\epsilon\over 2\pi\ell^2}$. 
This provides the 3D origin of the temperature \pdd\ obtained directly in 2D. 

In the usual extreme limit of BTZ, discussed in the end of the previous subsection, the 
coordinates are kept fixed as the temperature is lowered. Regular CFTs have a 
gap in the allowed conformal dimensions, so this extreme limit leaves just the ground 
state. The DLCQ limit scales the coordinates such that the periodicity increases. Then the 
energy of states are lowered along with the temperature is lowered and so a nontrivial 
sector remains. Alternatively, increasing the periodicity first we expect a continuous spectrum
for the non-compact theory, with states remaining at arbitrarily low temperature. 

In the DLCQ description the vacuum state encoded in the $L$ sector is notoriously complicated. 
We can discern the temperature of this state by comparing the identifications \sdc\ to the
more conventional \sdb. We find the effective 2D temperature
\eqn\sfz{
T^{\rm FT}_+ = {1\over 2\pi}~,
}
after compensating for the relative factor of $2$ between dimensionless temperatures in 
3D and 2D. 
This is the self-dual temperature, a strongly coupled value in the sense that
the CFT is not well approximated by a free gas. The temperature \sfz\  is the 
Frolov-Thorne temperature of the ground state, but this is not the sector we 
focus on in this paper. 

The physical energy is encoded in the stress tensor dual to AdS$_3$, projected along 
the direction that forms the boundary of AdS$_2$. We have
\eqn\dd{T_{tt}^{\rm 3D}= {4\epsilon^2\over \ell^2}\left[(r_+-r_-)^{-2}T_{--}^{\rm 
3D}+(r_++r_-)^{-2}T_{++}^{\rm 3D}\right]~,}
where we used the linear coordinate transformation \xdc\ and the definition 
of $\epsilon$ \xcc. The standard 3D expressions computed from \cf\ are
\eqn\ddb{T_{--}^{\rm 3D}={1\over 32\pi G_3L}(r_+-r_-)^{2}~,\quad T_{++}^{\rm 
3D}={1\over 32\pi G_3L}(r_++r_-)^{2}~,}
and so 
\eqn\de{\eqalign{T_{tt}^{\rm 3D}&={1\over 4\pi G_3L}{\epsilon^2\over \ell^2}}~.}
It is interesting that the two terms in \dd\ give the same contribution to the total energy
\de. In other words, the total 3D energy receives equal contributions from the excitations
(in the $R=-$ sector) and the inert condensate (in the $L=+$ sector). 


\newsec{Implications for Kerr/CFT}
At this point we have computed boundary currents, central charge and the asymptotic symmetry 
group of AdS$_2$ gravity directly in 2D, and we have validated our results by comparison
with a near DLCQ limit of 3D pure gravity. We can now apply the results to the thermodynamics of near extremal 
Kerr black holes.

First we review the thermodymics of the Kerr solution. When manipulating thermodynamic 
formulae it is useful to introduce the Planck length $l_P$ 
through
\eqn\ja{
G_4 = l^2_P ~.
}
In this notation, we are interested in near extremal Kerr black holes for which the mass
\eqn\jb{
M = {\sqrt{J}\over l_P} + E~,
}
with excitation energy $E \ll \sqrt{J}/l_P$ (with $\ell^2\sim2J l_P^2$ from \bd), 
or equivalently
$E\ell/J \ll 1$. The excitation energy is related to the non-extremality parameter
$\epsilon$ of the effective AdS$_2$ black hole as
\eqn\eac{\epsilon^2\lambda^2=4\sqrt{J}E\ell_P^3~,}
where $\lambda\ll  1$ is the scaling parameter defining the near extremal limit
(details in Appendix A). 
  
The black hole entropy for a general Kerr black hole is
\eqn\ja{\eqalign{
S & ={A\over 4G_4} =  2\pi (M^2 l^2_P + \sqrt{M^4 l^4_P - J^2})~.\cr
}}
For small excitation energy the Hawking temperature becomes
\eqn\jaa{\eqalign{
T_H &= {1\over 2\pi l_P} J^{-3/4}\sqrt{El_P}\cr &={\epsilon \lambda\over 2\pi \ell^2}~,}
}
so we can write the excitation energy as
\eqn\jab{
E = 4\pi^2 J^{3/2} T^2_H l_P~.
}
This is the same form as the high temperature behavior
\eqn\jac{
E = {\pi^2\over 6} cT^2_H R~,
}
of a general CFT on a circle of radius $R$. It is this
agreement in form that allows a description of the excitations in terms of a
dual CFT. 

We can understand some features of the dual CFT from our results in this paper. 
First, our result \pdc\ for the central charge becomes 
\eqn\eab{c={6\ell^2\over G_4} = 12J~,}
after inserting the AdS$_2$ radius \bd\ appropriate for the near horizon region of the
extreme Kerr black hole. This is the same value that Kerr/CFT assigns to the ground state,
but here it controls the excitations. We stress
again that our result \pdc\  for the central charge is intrinsic to 2D, although consistent with the 
Brown-Hennaux value \dce\ for AdS$_3$ through the map \pya\ to AdS$_2$. 

Comparison of \jab\ and \jac\ gives the scale $R$ of the dual CFT as
\eqn\jag{
R = 2\sqrt{J}l_P~.
}
This is slightly bigger than the AdS$_2$ radius $R=\sqrt{2}\ell$ and consistent
with the result reported in \CveticJN\foot{The model in \CveticJN\ allocates
half the energy to the zero-modes, as suggested
after \de. This doubles $R$ to $\sqrt{8}\ell$, while the energy and the
temperature of the excitations are $1/2$ of their values given in this section.}.

Now, the excitation energy $E$ and the temperature $T_H$ both vanish in the
strict extremal limit $\lambda\to 0$ as measured by asymptotic observers
using the time $t'$. However, we can rescale the physics
in the manner made explicit for the 3D/2D reduction in the previous section. In the Kerr
context we use $t=\lambda t'$ and the temperature  $T=T_H/\lambda$ is finite. The
conformal weight assigned to a state is  finite {\it after} the rescaling and the extremal limit  and
it reads
\eqn\jah{
h = {ER\over \lambda^2} = {c\over 24} \left( {2\pi T_H R\over\lambda}\right)^2
= {c\over 24} {2\epsilon^2\over\ell^2} ~.
}
This is the value that exactly reproduces the entropy of the excitations in \ja\ in the near extremal limit, {\it i.e.} 
\eqn\jao{S = 2\pi \left( {c_L\over 12} + \sqrt{c_Rh\over 6}\right) + \ldots~.}
with $c_{L,R}=c=12J$. After rescaling the conformal weights we have $h\gg c$ in the black hole regime $\epsilon\gg\ell$ so that Cardy's formula applies.

In the AdS$_2$ theory, we employed the scale $\ell$ to measure energies in \pd, but
this is not the scale describing the breaking of conformal symmetry in \jac. The appropriate comparison with the discussion in section 2.4 is energy per unit length
\eqn\jax{{E\over R \lambda}={\pi^2 \over 6}cT^2={T_{tt}+{\cal B}_t J_t\over \ell}~.}
The unambiguous point of comparison between the 4D geometry and AdS$_2$ is the
time coordinate $t$. This ensures the agreement in \jax, despite the ambiguity in the
scale relevant for $E$ in the 2D theory, due to the $SL(2,R)$ invariance of the theory.


\newsec{Discussion}
One of the highlights of this paper is the agreement between the entropy inferred
from the study of the conformal symmetry acting on the quantum theory, and the
entropy computed from the area law applied to the Kerr metric. As in many similar agreements
of the sort, a key step in the
computation is the presentation of the entropy from either point of view in the
form of a Cardy's formula. 

Now, in the context of the BTZ black hole it is an important point that the Cardy form
of the entropy, as well as the numerical values of its various parameters, are in fact
guaranteed by symmetries. The reported agreements are therefore {\it automatic}, in the
sense that they follow from symmetries (see eg. \KrausVZ). 
It is interesting to ask whether this sort of automatic agreement extends to the AdS$_2$ 
gravity considered in this paper.  

A good starting point is the free energy (on-shell action) of the $SL(2,R)$ invariant 
ground state ({\it ie} with $L_0=0$),
\eqn\ua{
I_{\rm gs} = \beta (L_0 - {c\over 24} )  = \beta( -{c\over 24} )~.
}
The black hole is the high temperature dual with free energy 
\eqn\ub{
I_{\rm bh} =(2\pi)^2 T ( -{c\over 24})~.
}
obtained from \ua\ by $2\pi\beta\to {1\over 2\pi\beta}$.
This latter expression is equivalent to Cardy's formula, upon transforming to the 
micro-canonical ensemble. 

The relation between ground state and highly excited modes in 2D CFT is of course 
very well known, but what we seek is the
corresponding relation on the gravitational side, which then guarantees the
accuracy of the accounting for the black hole entropy. In 3D the 
geometrical origin of the duality is the large coordinate
transformation exchanging the temporal and the azimuthal period in Euclidean gravity
(see {\it eg.} \KrausVZ). 
In particular, the azimuthal circle is contractible in global AdS$_3$ while the
temporal circle is contractible for the (non-rotating) BTZ black hole \DijkgraafFQ. 

In our discussion of AdS$_2$ quantum gravity the description as
an AdS$_2$ black hole applies for large $\epsilon^2/\ell^2\gg 1$, which ensures
$h_R\gg c$ such that the asymptotic form of the degeneracy applies. On the other
hand, the $SL(2,R)_R$ invariant vacuum state is global AdS$_2$, corresponding to
$\epsilon^2/\ell^2 = -1$. The high-low temperature duality of the CFT
relates these limits, and so provides the key ingredient 
for a gravitational justification of the Cardy formula, which in turn implies 
an automatic origin of the agreements. 

The aspect of the agreement that remains mysterious is the geometrical interpretation 
of the duality symmetry. Ultimately it should follow from modular invariance in 3D but
the details are nontrivial. For example the reduction is incompatible with the Euclidean 
continuation (see eg. \xdc), the customary setting for modular invariance. It 
would be nice to understand this feature in complete detail.

\bigskip
\noindent {\bf Acknowledgments:} \medskip \noindent   
We thank M. Cheng, T. Hartman, A. Lepage-Jutier, A. Maloney and A. Strominger for discussions. 
CK and FL thank the Aspen Centre for Physics for hospitality
during part of this work. The work of AC is supported in part by the National Science and Engineering Research Council of Canada. The work of CK is supported by the
Fundamental Laws Initiative at the Center for the Fundamental Laws of Nature. The work of FL is supported in part by the US DoE. 

\lref\KrausVZ{
  P.~Kraus and F.~Larsen,
  ``Microscopic Black Hole Entropy in Theories with Higher Derivatives,''
  JHEP {\bf 0509}, 034 (2005)
  [arXiv:hep-th/0506176].
}
\lref\DijkgraafFQ{
  R.~Dijkgraaf, J.~M.~Maldacena, G.~W.~Moore and E.~P.~Verlinde,
  ``A black hole farey tail,''
  arXiv:hep-th/0005003.
}


\appendix{A}{Near horizon geometry of the Kerr black hole}

The general Kerr solution is given by
\eqn\xa{\eqalign{ds^2=&-{\Sigma\Delta\over (r^2+a^2)^2-\Delta 
a^2\sin^2\theta}dt'^2+{\Sigma}\left[{dr^2\over\Delta}+d\theta^2\right]\cr &
+{\sin^2\theta\over\Sigma}((r^2+a^2)^2-\Delta a^2\sin^2\theta)\left[d\phi'-{2a\mu
r\over (r^2+a^2)^2-\Delta a^2\sin^2\theta}dt'\right]^2~,}}
with
\eqn\xab{\Delta=(r-r_-)(r-r_+)~,\quad r_\pm = \mu\pm \sqrt{\mu^2-a^2~,}}
and
\eqn\xac{\Sigma=r^2+a^2\cos^2\theta~.}
In our notation $\mu = G_4M$ and $a=J/M$ are length scales, while $J$ is 
dimensionless. The
near horizon region is isolated by introducing the coordinates
\eqn\xad{r={1\over2}(r_++r_-)+\lambda U~,\quad t'={t\over \lambda}~,\quad \phi' = 
\phi+{t\over \lambda (r_++r_-)}~.}
In the strict near horizon limit the dimensionless scaling parameter $\lambda\to 0$ , with $t, 
U, \theta, \phi$ fixed. In the extremal limit $r_+=r_-$, the metric is known as NHEK
geometry \BardeenPX. This limit is easily modified to maintain some energy above 
extremality. We need to take the limit while tuning the black hole parameters
such that the scale $\epsilon$ defined through
\eqn\xb{{1\over 2}{(r_+-r_-)}=\sqrt{\mu^2-a^2}\equiv \epsilon\lambda~,}
is kept fixed as $\lambda\to 0$. The resulting line element reads
\eqn\xbc{\eqalign{ds^2=&{1+\cos^2\theta\over2}\left[-{U^2-\epsilon^2\over 
\ell^2}dt^2+{\ell^2\over
U^2-\epsilon^2}dU^2+\ell^2d\theta^2\right]+\ell^2{2\sin^2\theta\over1+\cos^2\theta}\left(d\phi+{U\over\ell^2}dt\right)^2~,}}
with
\eqn\xaf{\ell^2 \equiv {1\over 2}(r_++r_-)^2=2\mu^2~,}
fixed. The $(t,U)$ term in the square brackets is
locally AdS$_2$ with radius of curvature $\ell$ but the global structure is modified 
into a black
hole geometry with horizon located at  $U=\epsilon$.  The near extreme limit described 
here is the
same considered in \refs{\AmselEV,\DiasEX} (and a Kerr analogue
of  ``Limit 2'' in \MaldacenaUZ).


\appendix{B}{Fefferman-Graham expansion in 2D}

In this appendix we provide further details about the  Fefferman-Graham expansion in 
2D and some useful formulas.  We write the metric and gauge fields as
\eqn\ka{\eqalign{h_{tt}&=h^{(0)}e^{2\rho/\ell}+h^{(2)}+h^{(4)}e^{-2\rho/\ell}+\cdots~,\cr{\cal 
B}_{t}&={\cal B}^{(0)}e^{\rho/\ell}+{\cal B}^{(2)}+{\cal
B}^{(4)}e^{-\rho/\ell}+\cdots~, }}
and take $\psi=0$. We can relate the coefficients of the asymptotic expansion by using Maxwell's eqn. This tells us that the field strength is proportional to the volume form
 \eqn\kb{d{\cal B}=\ell\,\epsilon_{{\rm AdS}_2}~,}
and leads to
\eqn\kc{\eqalign{ h^{(0)}&=-\ell^2{\cal B}^{(0)}{\cal B}^{(0)}~,\cr h^{(2)}&=2\ell^2 
{\cal B}^{(0)}{\cal B}^{(4)}~,\cr h^{(4)}&=-\ell^2\left({\cal B}^{(4)}{\cal
B}^{(4)}-6{\cal B}^{(0)}{\cal B}^{(8)}\right)~,\cr h^{(6)}&=\ell^2\left(10 {\cal
B}^{(0)}{\cal B}^{(12)}-6 {\cal B}^{(4)}{\cal B}^{(8)}\right)~.}  }

For the black hole solution \bc, \bca\ the non-zero coefficients of the expansion are
\eqn\kd{h^{(0)}=-{1\over 4}~, \quad h^{(2)}={\epsilon^2\over 
2\ell^2}~, \quad h^{(4)}=-{\epsilon^4\over 4\ell^4}~,}
and
\eqn\kda{{\cal B}^{(0)}={1\over 2\ell}~, \quad {\cal B}^{(2)}=-{\epsilon\over 
\ell^2}~, \quad {\cal B}^{(4)}={\epsilon^2\over 2\ell^3}~.}

The 3D boundary stress tensor is given by \cf, with the extrinsic curvature defined as
\eqn\kga{\eqalign{K_{ab}&={1\over2L}\partial_\eta\gamma_{ab} \cr &={1\over 
L}e^{2\eta}\gamma_{ab}^{(0)}-{1\over L}e^{-2\eta}\gamma_{ab}^{(4)} +\cdots ~,}}
and its trace $K=\gamma^{ab}K_{ab}$, where
\eqn\kg{\gamma^{ab}=e^{-2\eta}\gamma^{(0)ab}-e^{-4\eta}\gamma_{cd}^{(2)}\gamma^{(0)ac}\gamma^{(0)bd}+\cdots 
~.}
Using \kga\ and \kg, the stress tensor \cf\ in terms of the boundary metric \hcd\ is
\eqn\kgb{\eqalign{T^{\rm 3D}_{ab}&={1 \over 8\pi G_3 L}\left[\gamma_{ab}^{(2)}-\gamma_{ab}^{(0)}{\rm Tr}(\gamma^{(2)}) +e^{-2\eta}\left(2\gamma_{ab}^{(4)}-\gamma_{ab}^{(2)}{\rm Tr}(\gamma^{(2)})-\gamma_{ab}^{(0)}{\rm Tr}(\gamma^{(4)})\right)+\cdots\right]~,}}
with ${\rm Tr}(\gamma^{(n)})=\gamma^{(0)ab}\gamma_{ab}^{(n)}$.  In terms of the 2D fields \ka\ the boundary metric is given by \hdb, and  the 
components of \kgb\ reduce to
\eqn\kf{\eqalign{T^{\rm 3D}_{tt}&={\ell^2 \over 8\pi G_3 L}\left[{\cal B}^{(2)}{\cal 
B}^{(2)}+4{\cal B}^{(0)}{\cal B}^{(4)}+12e^{-2\eta}{\cal B}^{(2)}{\cal B}^{(4)}+{\cal
O}(e^{-4\eta})\right]~,\cr
T^{\rm 3D}_{t\theta}&={\ell^2 \over 8\pi G_3 L}\left[{\cal B}^{(2)}+6e^{-2\eta}{\cal
B}^{(4)}+{\cal O}(e^{-4\eta})\right]~,\cr
T^{\rm 3D}_{\theta\theta}&={\ell^2 \over 8\pi G_3 L}\left[1+4e^{-4\eta}{{\cal
B}^{(4)}\over {\cal B}^{(0)}}+{\cal O}(e^{-6\eta})\right]~,}}
where we used \kc\ to write $h^{(n)}$ as a function of ${\cal B}^{(n)}$.

 \listrefs
\end